%% file: astroph.tex
\documentclass[10pt]{article}
\usepackage{epsfig,rotating}

\def\RNn{\hbox{$n$(\NII)/$n$(\NI)}}
\def\RNN{\hbox{$N$(\NII)/$N$(\NI)}}

\def\ebv{\hbox{$E(B-V)$}}
\def\kms{\hbox{km s$^{-1}$}}
\def\deeg{\hbox{$^\circ$}}
\def\glong{\hbox{$\ell$}}
\def\glat{\hbox{$b$}}
\def\cmtwo{\hbox{cm$^{-2}$}}
\def\NHI{\hbox{$N$(H$^{\rm o }$)}}
\def\logNHI{\hbox{log$N$(H$^{\rm o }$)}}
\def\logNH{\hbox{log$N$(H)}}
\def\logNH{\hbox{log$N$(H)}}

\def\nHI{\hbox{$n$(H$^\circ$)}}
\def\nHIavg{\hbox{$\langle n$(H$^\circ$)$\rangle$}}
\def\nHII{\hbox{$n$(H$^+$)}}
\newcommand{\Htwo}{H\,\textsc{II}}
\def\nHeI{\hbox{$n$(He$^\circ$)}}
\def\nel{\hbox{$n$(e)}}
\def\np{\hbox{H$^+$}}
\def\HeI{\hbox{He$^\circ$}}
\def\NeI{\hbox{Ne$^\circ$}}
\def\OI{\hbox{O$^\circ$}}
\def\NI{\hbox{N$^\circ$}}
\def\NII{\hbox{N$^+$}}
\def\NHH2{\hbox{$N$(${\rm H^{ \rm o } + 2H_2}$)}}

\def\NDI{\hbox{$N$(D$^{\rm o }$)}}
\def\HI{\hbox{H$^{ \rm o }$}}
\def\HII{\hbox{H$^{ \rm + }$}}
\def\DI{\hbox{D$^{ \rm o }$}}
\def\MgI{\hbox{Mg$^{ \rm o }$} }
\def\MgII{\hbox{Mg$^{ \rm + }$}}
\def\CI{\hbox{C$^{ \rm 0 }$}}
\def\CIIstar{\hbox{C$^{ \rm +* }$}}
\def\CII{\hbox{C$^{ \rm + }$}}
\def\SiIII{\hbox{Si$^{ \rm ++ }$}}
\def\NCaII{\hbox{$N$(Ca$^+$)}}
\def\CaII{\hbox{Ca$^+$}}
\def\NaI{\hbox{Na$^o$}}
\def\CaIII{\hbox{Ca$^{++}$}}
\def\CaIII{\hbox{Ca$^{++}$}}
\def\cc{\hbox{cm$^{-3}$}}
\def\Pmag{\hbox{P$_\mathrm{B}$}}
\def\Ptherm{\hbox{P$_\mathrm{TH}$}}

\def\muG{\hbox{$\mu$G}}
\input{6001-journallist}
\begin{document}
\begin{center}
\begin{Large}{ \bf The Sun's Journey Through the Local Interstellar Medium: The
PaleoLISM and Paleoheliosphere} \\
\end{Large}
\vspace{0.15in}
\normalsize
P.~C.~Frisch
\footnote{University of Chicago, Department of Astronomy and Astrophysics, 5640 S. Ellis Ave., Chicago, IL  60637, U.S.A.. Correspondence:  P. Frisch (frisch@oddjob.uchicago.edu)}
 and J.~D.~Slavin
\footnote{Harvard-Smithsonian Center for Astrophysics, 60 Garden Street,
Cambridge, MA 02138}
\end{center}

\vspace{0.2in}
\begin{center}Abstract\end{center} 
Over the recent past, the galactic environment of the Sun has differed
substantially from today.  Sometime within the past $\sim$130,000
years, and possibly as recent as $\sim$56,000 years ago, the Sun
entered the tenuous tepid partially ionized interstellar material now flowing
past the Sun.  Prior to that, the Sun was in the low density interior
of the Local Bubble.  As the Sun entered the local ISM flow, we passed
briefly through an interface region of some type.  The low column
densities of the cloud now surrounding the solar system indicate that
heliosphere boundary conditions will vary from opacity considerations
alone as the Sun moves through the cloud. These variations in the
interstellar material surrounding the Sun affected the
paleoheliosphere. 
\section{Introduction}

The boundary conditions of the heliosphere at any given point in its
history are set by the interstellar cloud that happened to surround
the Solar System at that time.  Our variable galactic environment
affects the physical properties of the heliosphere, and the fluxes of
galactic cosmic rays and interstellar byproducts reaching the Earth.
The work of Hans Fahr has, from the earliest until now, provided a
strong foundation for our understanding of the effect of the ISM on
the heliosphere and on Earth
\cite{Fahr:1968,Fahr:1974,RipkenFahr:1983,SchererFahr:2003,YeghikyanFahr:2004}.

The properties of both astrospheres and the heliosphere are highly
responsive to the boundary conditions supplied by the interstellar
medium and the interstellar radiation field
\cite{Fahr:1978,Frisch:1993,Frisch:1997,ZankFrisch:1999,Schereretal:2002,Florinskietal:2003,Muelleretal:2006,Frisch:2006}.
The space motion of the Sun, when compared to the column densities of
interstellar material (ISM) towards stars near the Sun, indicates that
sometime in the late Quaternary the Sun, which has been moving through
the very low density region known as the Local Bubble, encountered the
cluster of local interstellar clouds (CLIC) flowing away from the
direction of the Scorpius-Centaurus Association
\cite{Frisch:1981,Frisch:1997,FrischYork:1986,FrischSlavin:2006}.
Mediating the interaction between the very low density Local Bubble
and the tepid CLIC will be a thin interface of some type
\cite{Slavin:1989,SlavinFrisch:2002}.

Regardless of the Local Bubble plasma pressure, the large contrast
($>10^3$) between neutral ISM densities in the Local Bubble versus the
CLIC must have generated significant changes in the heliosphere,
because pickup ions and anomalous cosmic rays are processed
interstellar neutrals.  The well-known anticorrelation between
galactic cosmic ray fluxes at Earth and the solar magnetic activity
cycle, mediated by the heliosphere, suggests that the transition
between the Local Bubble cavity and CLIC altered both the heliosphere
and the galactic cosmic ray flux at Earth, with possible implications
for the terrestrial climate.  Heliosphere models have now been
constructed for a range of interstellar boundary conditions
\cite{Muelleretal:2006,FlorinskiAxford:2003,Scherer:2000,YeghikyanFahr:2004}.
The models show that the ISM surrounding the Sun during the geological
past (or the ``paleoLISM'') did affect the heliosphere during the
geological past (or the ``paleoheliosphere''), and therefore also the
galactic cosmic ray flux at Earth.

In this paper, we postulate a geometric model for the CLIC,
supplemented by equilibrium models that provide the cloud density.
The entry epoch of the Sun into the CLIC (also known as the Local
Fluff) depends on the relative Sun-CLIC velocities and the
distribution of the CLIC gas.  To estimate this transition epoch, the
distribution of CLIC gas is derived from a simple model of the 
cloud morphology using data for \HI\ and \DI\ towards nearby stars (\S
\ref{sec:distribution}), combined with  the CLIC density found from 
photoionization equilibrium models of nearby ISM (\S \ref{sec:rt}).
The entry of the Sun into the CLIC is then found from the cloudlet
velocities in the downwind direction (\S \ref{sec:distribution}).
Prior to encountering the CLIC, the Sun was in the very low density
gas of the Local Bubble interior for several million years (\S
\ref{sec:localbubble}).  The limitations of the simple assumptions
 underlying these estimates are discussed briefly in \S
 \ref{sec:assumptions}.

\section{Sun Passage into Very Local ISM} 

The epoch when the Sun first encountered the CLIC gas can be estimated
from data on ISM column densities and velocities towards nearby stars,
combined with theoretical models that provide the average cloud
density.  These photoionization models provide a
second important result about the paleoheliosphere, by showing clearly
that the boundary conditions of the heliosphere, in particular the ionization
level of hydrogen and the electron density, vary from radiative
transfer effects alone as the Sun traverses low opacity ISM (\S
\ref{sec:rt}, \cite{SlavinFrisch:2002,SlavinFrisch:2006}, SF02, SF06).

\subsection{Local ISM Data and Distribution } \label{sec:distribution}

The Sun has recently entered the CLIC, which itself is inhomogeneous.
To estimate the entry date, we map the CLIC gas by assuming that the
distances to the CLIC edges are given by \NHI/\nHIavg, where \NHI\
are column densites towards nearby stars.  The
average space density \nHIavg\ is provided by radiative transfer models
(\S \ref{sec:rt}).

CLIC dynamics indicate an origin
related to a superbubble caused by star evolution in the
Scorpius-Ophiuchus Association \cite{Frisch:1981}.  The bulk flow
velocity vector of the CLIC gas past the Sun is 
$\sim$--28$\pm$4.6 \kms, from the direction
\glong,\glat$\sim$12\deeg,12\deeg\ in heliocentric coordinates
\cite{FGW:2002}.  This vector is based on absorption
components towards $\sim$60 stars, obtained at resolutions of 0.3--3.0
\kms.  Most of the Hyades stars were excluded from the star sample
underlying this vector, because ISM with a poorly defined relationship
to the CLIC is found inside of this cluster \cite{RLhyades}.  The LIC
is the best understood member of the CLIC, with a precisely known
heliocentric velocity based on Ulysses \HeI\ data of $-26.3\pm0.4$ \kms\ from
\glong,\glat = 3.3\deeg,15.9\deeg\ \cite{Witte:2004}.  The
exact value for the CLIC bulk flow depends on the underlying star
sample, since the flow gradually decelerates towards the downwind
direction (FS06).  Examples of the deceleration are the LIC, found in the
downwind direction at +26 \kms, and the Apex cloud, found at --35
\kms\ in the upwind direction (HC).

The velocity vectors of the CLIC, LIC, and the Sun are listed in Table
\ref{tab:1} in both HC coordinates and the local standard of rest
(LSR) velocity frame, for both the Standard and Hipparcos-based LSR
frame.  The uncertainty in the LSR occurs because the age
distributions of the star samples underlying these two LSR vectors
were different \cite{DehnenBinney:1998,Mihalas:1981}.

Data on \NHI\ are drawn from observations of \DI\ and \HI\ towards
nearby cool and hot stars \cite{Woodetal:2005,RLII,FGW:2002}, white
dwarf stars \cite{Wolffetal:1999,Lehneretal:2003,Vallerga:1996,
Oliveiraetal:2003,HebrardMoos:2003,Woodetal:2002wd,Lemoineetal:2002,
Kruketal:2002,Frisch:1995}, and nearby stars with observations of
interstellar \CaII\ (\cite{FGW:2002,CrawfordLallementWelsh:1998,CrawfordDunkin:1995}, and references therein).  The
column densities of all velocity components towards a star are summed
to obtain the total \NHI\ towards the CLIC surface.  For stars with
\DI\ data, a conversion factor of \NHI/\NDI=1.5$\times 10^{-5}$ is
used.  The total \HI\ column densities to the CLIC surface are in the
range \NHI$ = 0.3 \times 10^{18}$ to $\sim 10^{19}$ \cmtwo.

Optical \CaII\ data are also used, however the ratio \NCaII/\NHI\ is
highly variable.  In cold clouds $\sim$99.7\% of the Ca is depleted
onto dust grains.  Both depletion and ionization affect \CaII\ in warm
clouds such as the CLIC; for example \CaIII/\CaII$>$1 if $T>4,000$ K
and \nel$<$0.13 \cc\ \cite{WeltyCa:1996}.  The ratio
\NHI/\NCaII=$ 10^{-8}$ is used, based on the three absorption
components observed in both \CaII\ and \HI\ towards $\alpha$ Aql
($d$=5 pc, \glong,\glat$\sim$48\deeg,--9\deeg, e.g., \cite{FGW:2002,RLII,Ferlet:1986}).

Looking only at stars within 10 pc, average values of
$\langle$\NHI$\rangle \sim 10^{18}$ \cmtwo\ and $\langle$\nHI$\rangle
\sim 0.07$ \cc\ are found.  Based on temperature and turbulence data
in Redfield and Linsky (2004), objects within 10 pc show a temperature and
turbulence range of $T$=1,700--12,600 K and $\xi=0-5.5$ \kms, with
mean values $6,740 \pm 2800$ K and $\xi =1.9 \pm 1.0$ \kms.  The $\xi$
variable represents deviations from a Maxwellian velocity distribution 
for the atoms contributing to the absorption line components, and as 
such is a mock turbulence that includes unresolved clouds.

The range of temperatures inferred for the CLIC,
combined with the macroturbulence of $\pm$4.6 \kms,
show that the CLIC is inhomogeneous and that the boundary conditions of the heliosphere will vary
during the next thousands of years as the Sun traverses the CLIC.

\begin{figure}[t]
  \vspace*{2mm}
  \begin{center}
\includegraphics*[width=3.2in]{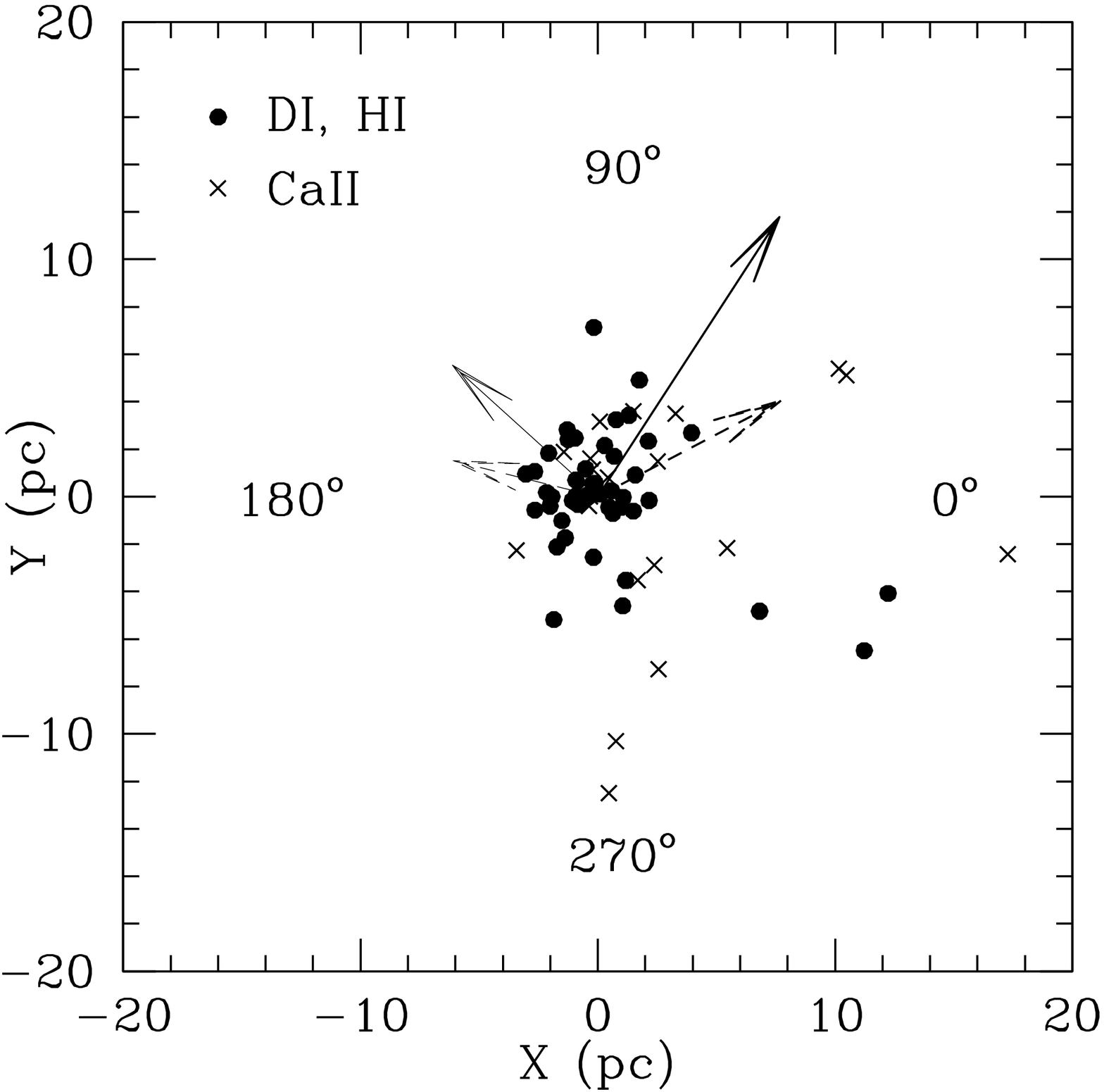}
  \end{center}
  \caption{The distances to the edge of the CLIC, projected
onto the galactic plane, as calculated
from \NHI/\nHIavg\ for stars near the Sun.  Distances are based on 
observations of \HI\ (dots), \DI\ (dots), and \CaII\ (crosses) 
(see text).  The directions \glong=0\deeg, 90\deeg,
180\deeg, and 270\deeg\ are labeled.
The arrows directed to the left show the CLIC motion through the Local
Standard of Rest (LSR) based on the standard (solid) and Hipparcos
(dashed) solar apex motions, while the arrows to the right show the
motion of the Sun through the LSR based on the Hipparcos and Standard
apex values (Table \ref{tab:1}).}
  \label{fig:clicxy}
\end{figure}

\begin{figure}[t]
  \vspace*{2mm}
  \begin{center}
\includegraphics*[width=3.2in]{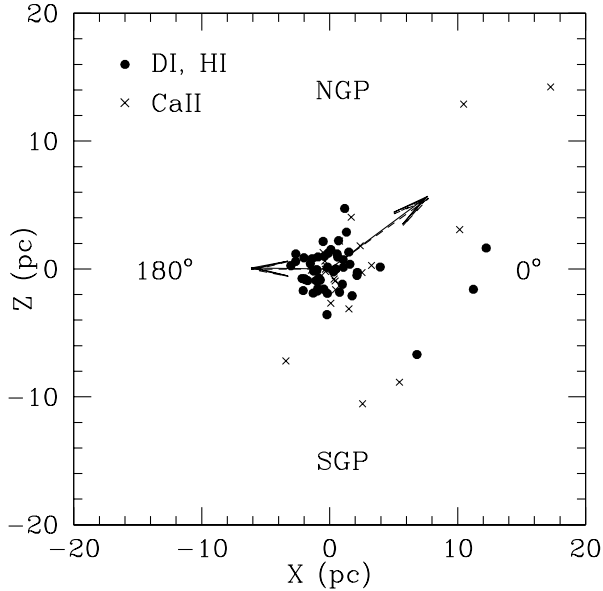}
  \end{center}
  \caption{Same as Fig. \ref{fig:clicxy}, except that the projection 
is on a meridian plane perpendicular to the galactic plane.  }
  \label{fig:clicxz}
\end{figure}

The CLIC is fully contained in the nearest $\sim$35 pc of space.  The
distances to the edges of the CLIC are shown in Figs. \ref{fig:clicxy}
and \ref{fig:clicxz}. These figures are constructed from the data
listed above, and assuming $\langle$\nHI$\rangle$=0.17 \cc\ (discussed
in \S \ref{sec:rt}). This ISM distribution is assumed to contain no
gaps.  


An extended dilute \Htwo\ region found towards $\lambda$ Sco
\cite{York:1983} may contribute to the excess cloud length towards
the galactic center that is indicated by \CaII\ data.  Highly ionized
gas at CLIC velocities is found towards both $\lambda$ Sco
(\glong,\glat=352\deeg,--2\deeg) and HD 149499B ($d$=37 pc,
\glong,\glat=330\deeg,--7\deeg), where for instance \NII/\NI=1.9
\cite{Lehneretal:2003}.

\subsection{Boundary Condition Variations from Cloud Opacity} \label{sec:rt}

An important influence on the heliosphere while the Sun traverses the
low opacity CLIC is the change in heliospheric boundary conditions
caused by ionization variations due to the attenuation of photons with
energy $E>$13.6 eV (the ionization edge of \HI).  These variations are
shown by our radiative transfer (RT) models
\cite{SlavinFrisch:2002}\footnote{By ``radiative transfer effects'',
we mean that the spectrum of the radiation field, including both point
source and diffuse contributions, is substantially modified for
energies $E>13.6$ eV as the radiation penetrates more deeply into the
cloud.  This is the case for the CLIC gas, as shown by the relative
opacities of the cloud to H- versus He- ionizing photons.}.  The CLIC
is partially opaque to H-ionizing photons ($\lambda < 912$ \AA) and
nearly transparent to He-ionizing photons ($\lambda < 504 $ \AA).  A
cloud optical depth of $\tau \sim$1 is achieved for
\logNHI$ \sim 17.2$
\cmtwo\ and $\sim 17.7 $ \cmtwo, respectively, at the \HI\ and \HeI\
ionization edges.

\begin{figure}[t]
  \vspace*{2mm}
  \begin{center}
   \includegraphics*[width=3.2in]{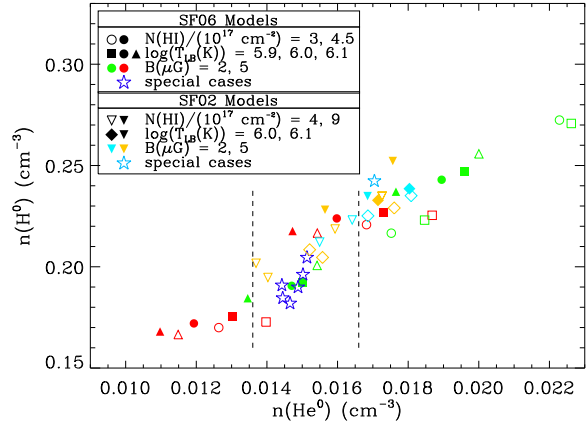}
  \end{center}
  \caption{
The densities of \HI\ and \HeI, as predicted by radiative transfer
models of the CLIC, are shown for a range of equilibrium conditions 
based on models in SF02 and SF06.  The symbols give the model parameters:
symbol colors indicate the magnetic 
field strength in the cloud interface; symbol fill gives gives the 
assumed HI column density; symbol shape gives the assumed temperature 
of the Local Bubble plasma.  The stars are special parameter sets
which do not fall on the grid of model parameters, but rather are chosen to
better match the data.  The range of current values for \nHeI, based on 
Ulysses and pickup ion data, are 
shown as vertical lines \cite{Moebiusetal:2004}.  Although a range of 
\nHI\ values are consistent with \nHeI, the 
radiative transfer models providing the best agreement with ISM
observations both inside of the heliosphere and towards nearby stars
give \nHI$=0.18- 0.21$ \cc.
}
\label{fig:HIvHeI}
\end{figure}

A series of models have been constructed to study these opacity
effects (SF02, SF06).  These models employ the CLOUDY radiative
transfer code, which incorporates a wide range of physical processes
to model a cloud under the conditions of photoionization equilibrium
\cite{Ferland:1998}. The models are constrained by 
observations of the CLIC towards $\epsilon$ CMa
\cite{GryJenkins:2001}, observations of pickup ions, anomalous cosmic
rays, and \nHeI\ inside of the solar system, and interstellar
radiation field data and models.  The radiation field is based
on radiation sources affecting the solar environment (see SF02),
and extends out to soft X-ray energies (\S \ref{sec:lbplasma}.
Fig. \ref{fig:HIvHeI} summarizes the variations in neutral densities
that are obtained for equilibrium calculations of low density ISM
similar to the LIC (see SF02 and SF06).  The best of these models
(models 2 and 8 in SF02) give
\nHI=0.18--0.21 \cc\ and \nHII=0.1 \cc\ at the solar location.  These
model results that best predict the observed gas densities, for a
study still in progress, give a mean value of $\langle$\nHI$\rangle
\sim$0.17 \cc\ for the sightline towards the downwind cloud surface.


\begin{figure}[t]
  \vspace*{2mm}
  \begin{center}
   \includegraphics*[width=3.2in]{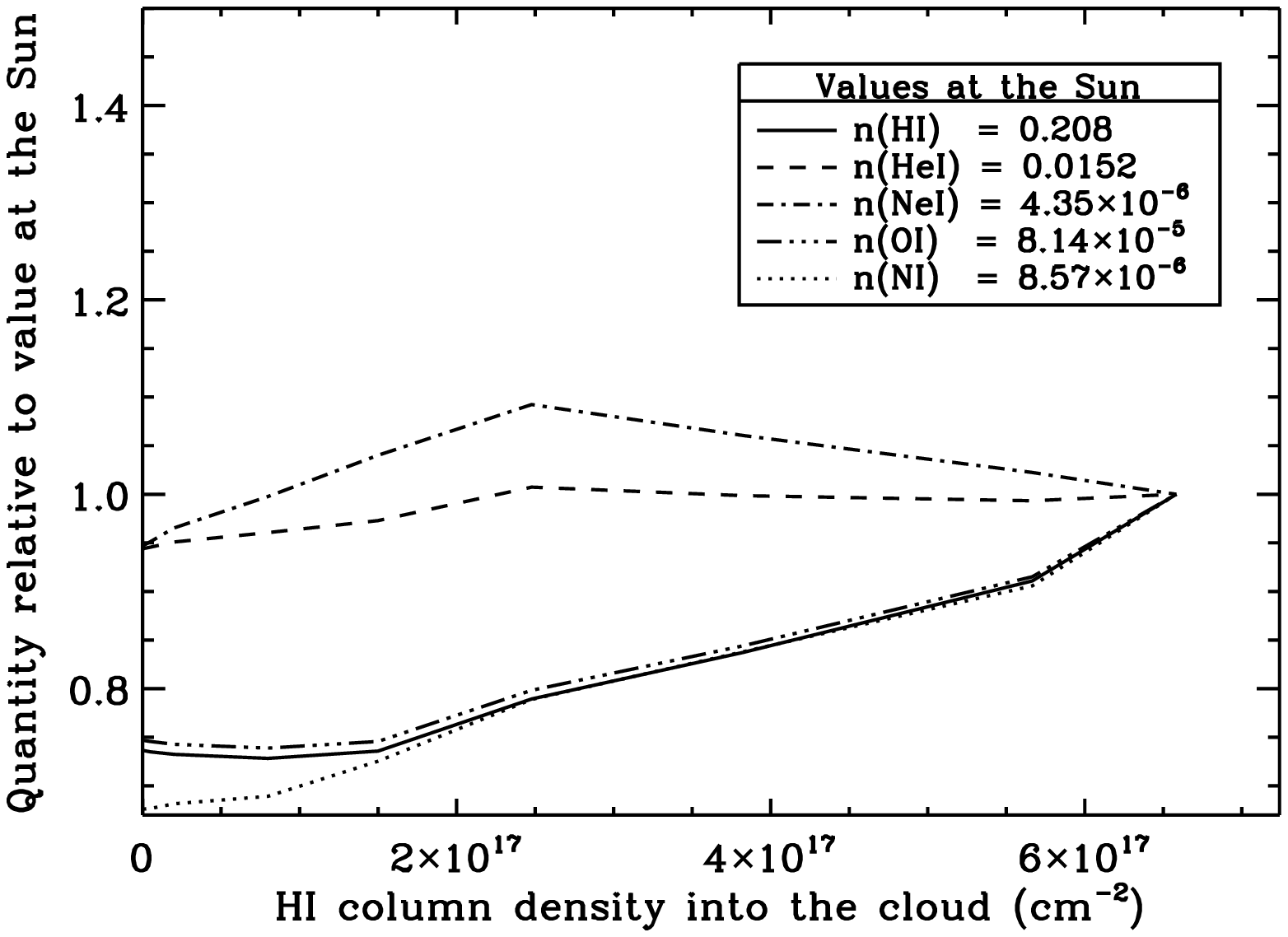}
  \end{center}
  \caption{
Variation of neutral densities, due to radiative transfer effects
between the Sun and cloud surface, for 
Model 2 from SF02.  Shown are
variations in neutral densities between the Sun (\NHI
=6.5$\times$ 10$^{17}$ \cmtwo) and cloud surface (\NHI=0) for \HI,
\HeI, \NeI, \OI, and \NI.  At the heliopause, \nHeI$\sim$0.015 \cc,
\nHI$\sim$0.21 \cc, \nel$\sim$0.1 \cc.  The cloud surface is at the
left, and the solar location is at the right of the figure.
}
  \label{fig:rt}
\end{figure}

Fig. \ref{fig:rt} shows the effects of cloud opacity.  The ionization levels
of H, O, and N, which are coupled by charge exchange, decrease by $\ge$20\%
from the cloud exterior to the solar location.  \footnote{The small decrease
in \NI/\HI\ at cloud edge results from a high \NI\ photoionization
cross-section.}  In contrast, He and Ne ionizations, which require photons
more energetic by $>$50\%, vary little.  Guesstimates indicate that
for $\sim$50\% filtration of \HI, 
converting 20\% of the H from \HI\ to \HII\ would
raise the H pressure confining the paleoheliosphere at the cloud surface 
on the order of $\sim$10\%, as compared to the present-day value.
The timescales over which these variations occur depend 
on the assumed cloud shape, and may be as short as thousands
of years.

Ionization variations that are five times larger than those we have
modeled are observed in the very local ISM.  Warm partially ionized
material (WPIM, $T>5000$ K) such as the CLIC is widespread near the
Sun, and is sampled by data on \NII\ and \NI\ from $Copernicus$ and
FUSE.  Values of \RNN$\sim$0.4--2.0 are typical for low density
sightlines, \NHI$< 10 ^{19}$ \cmtwo\
\cite{RogersonIII:1973,Oliveiraetal:2003,Lehneretal:2003}.
The best CLIC models give \RNn$\sim$0.76 at the Sun (nos. 2 and 8 in
SF02).  Consequently, as the CLIC sweeps past the Sun variations
larger than 10\% in the heliosphere may be caused by variable
interstellar ionizations.  The \HII\ gas towards $\lambda$ Sco and HD
149499B, 37 pc away in the upwind direction, is an example of an
ionized cloud that could engulf the Sun in the next million years.

\subsection{Solar Encounter with the CLIC} \label{sec:epoch}

When did the Sun enter the CLIC?  With the exception of the LIC, only the
Doppler-shifted radial components of cloud motions are observed.  Hence only
approximate estimates of the epoch that the Sun made the transition from the
very low density Local Bubble to the CLIC are possible.  We get an
averaged value for the time of encounter by calculating the distance to the
CLIC edge for all stars within 50 pc, and between
\glong=170\deeg$\pm$30\deeg\ and $|$\glat$| < $30\deeg, and comparing
this distance and the HC ISM velocity towards each star.  The distance
to the cloud edge is given by \NHI/\nHIavg, where \nHIavg=0.17 \cc.
Nine stars in Figs. \ref{fig:clicxy} and \ref{fig:clicxz} fall in this
interval.  The average distance to the cloud edges for these stars is
2.8 pc (range 1.2--3.8 pc), and the average entry time of the Sun into
the CLIC is 120,000 years ago (range 56,000--200,000 years ago).
Restricting the downwind stars to those within 30 parsecs yields
four stars, and does not significantly change these entry time values.  The
star closest to the HC downwind direction of the LIC
(\glong,\glat$\sim$183\deeg,--16\deeg) is $\chi^1$ Ori (HD 39587, at
8.7 pc).  The HC cloud velocity of 21.6 \kms\ towards this star
\cite{RLII} indicates an entry date into the CLIC of $\sim$56,000
years ago.  We conclude that for no gaps in the ISM distribution
within the CLIC, so that \nHIavg$\sim$0.17 \cc\ is valid, then the Sun
entered the CLIC within the past $\sim 130,000 \pm 70,000$ years, and
possibly within the past 56,000 years (consistent with
our earlier estimates, \cite{Frisch:1997,FrischSlavin:2006}).

\begin{figure}[h]
  \vspace*{2mm}
  \begin{center}
\includegraphics*[width=3.2in]{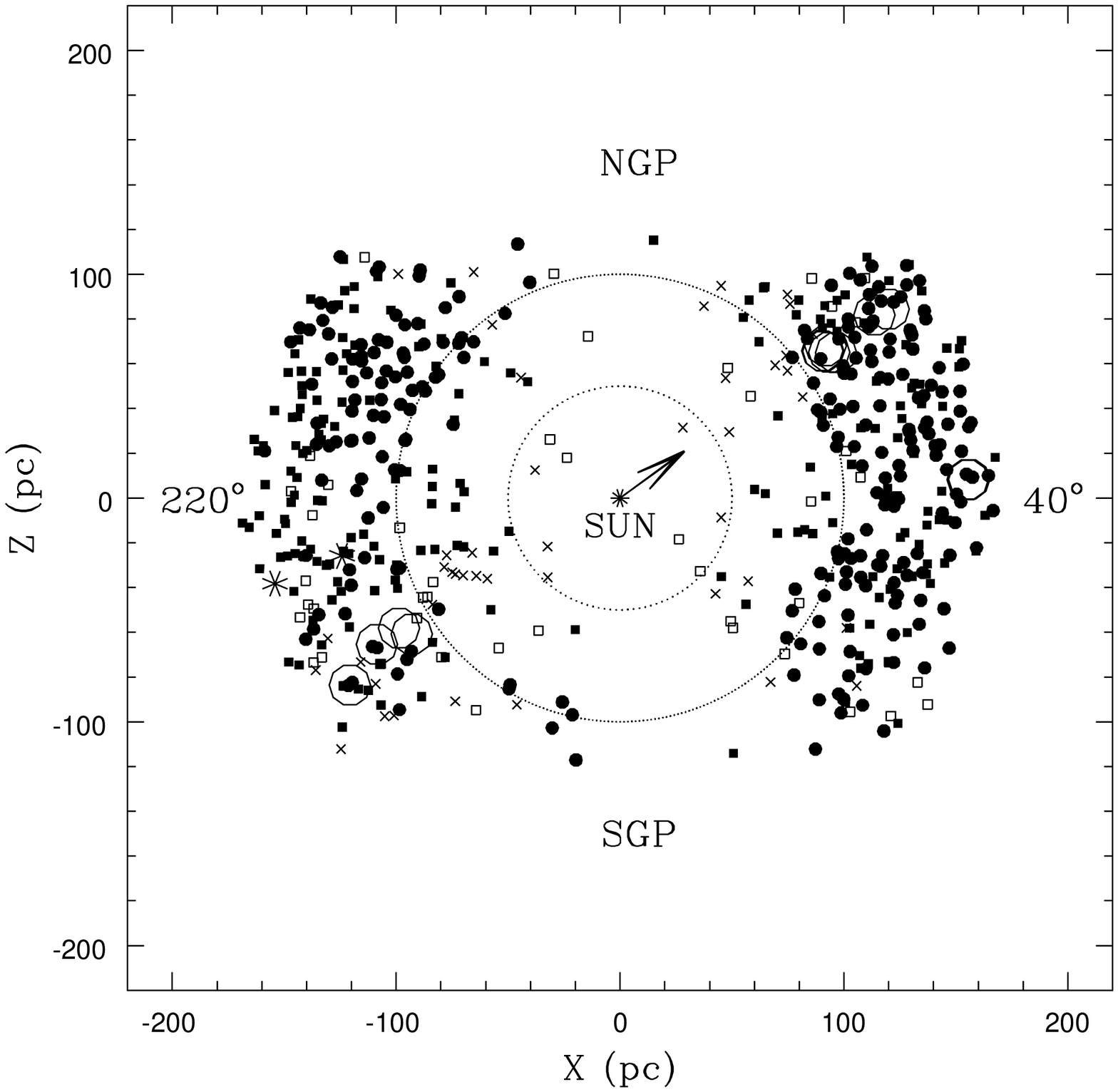}
  \end{center}

  \caption{The distribution of ISM is shown for non-variable stars within
$\pm$25\deeg\ of a meridian slice perpendicular to the galactic plane
and aligned with the solar apex motion through the LSR.  The solar apex motion
is plotted as an arrow.  The X-axis points towards a galactic longitude
of \glong=40$^\circ$, and the Z-axis points towards the North Galactic
Pole.  The filled symbols show sightlines with \ebv$>$0.06 mag (or
\logNH$>$20.48, dex \cmtwo), and the x's show sightlines with
\ebv$<$0.02 mag (or \logNH$<$19.93, dex \cmtwo).  Open squares have
\ebv=0.02--0.06 mag.  Large open circles display the positions of
nearby dust clouds (see text).  These reddening values are determined
from photometric data in the Hipparcos catalog
\cite{Perrymanetal:1997}.  Data points are smoothed over
$\pm$13\deeg, and incorporate all stars with overlapping distances
(once uncertainties are included).  The asterisks indicate the
positions of the stars $\epsilon$ CMa and $\beta$ CMa, which are
located in a well-known spatial region with low interstellar densities
\cite{FrischYork:1983}, but which are not included in the plotted
star sample because the Hipparcos catalog flags them as variable.  }
  \label{fig:lbxz}
\end{figure}

If the velocity dispersion of the CLIC clouds represents turbulence,
then the assumption of no gaps implies incompressible turbulence.  A
magnetized partially ionized tenuous plasma will have incompressible
turbulence if the magnetic pressure, \Pmag, is on the order of, or
greater than, the thermal pressure, \Ptherm.  Magnetic field strengths
of $B \sim$3 \muG\ are required in the LIC to balance thermal
pressures of \Ptherm$\sim$2500 \cc\ K, as expected from the best
photoionization models with \nel$\sim$0.1 \cc, \np$\sim$0.1 \cc,
\nHI$\sim$0.2 \cc, and T=6300 K.  Weaker fields would indicate
compressible turbulence and possibly gaps between the cloudlets.  In
such a case, the entry of the Sun into the CLIC may have occurred
earlier than stated above.

The LIC is observed towards both $\alpha$ CMa, at 2.7 pc, and
$\epsilon$ CMa (towards \glong$\sim$230\deeg,
\glat$\sim$--10\deeg, \cite{Hebrardetal:1999,GryJenkins:2001}).  The LIC
\NHI\ towards $\alpha$ CMa has been used to set constraints on the
entry date of the Sun into the LIC
\cite{Frisch:1994,Muelleretal:2006,FrischSlavin:2006}.  Although the
LIC velocity is well known, the cloud structure is uncertain, and we
can only state that the Sun first encountered the LIC within the past
40,000 years, and possibly within the past 3,000 years, based on the
earlier discussions.

\section{The Sun in the Local Bubble Cavity} \label{sec:localbubble}

Prior to crossing paths with the CLIC, the Sun traveled through the
Local Bubble (LB) for several million years \cite{FrischYork:1986}.
The path of the Sun through the Local Bubble is reconstructed below
from reddening data (\S \ref{sec:lbdata}).  The LB interior formed one
type of paleoLISM, and the limits on the plasma properties of the
Local Bubble interior are discussed in \S \ref{sec:lbplasma}.

\subsection{Local Bubble } \label{sec:lbdata}

The exact dimensions and structure of the Local Bubble depend on the
ISM component that is sampled; we use the optical reddening properties
of interstellar dust grains here, based on photometric data in the
Hipparcos catalog \cite{Perrymanetal:1997}, and find a radius of
$\sim$ 60--100 pc.  Earlier maps of the Local Bubble ISM distribution
based on reddening data that traces interstellar dust
\cite{Lucke:1978,Perry300:1982,PerryChristodoulou:1996,Vergelyetal:1997},
dust and magnetic fields traced by polarization data
\cite{MathewsonFord:1970,Leroy:1999}, and gas traced by \HI\ and
\NaI\ \cite{FrischYork:1983,Paresce:1984,Vergelyetal:2001,Lallementetal:2003}
yield similar conclusions about the bubble dimensions, although
details of the cavity topology depend on data sensitivity to low
column density ISM.  Fig.\ \ref{fig:lbxz} shows the motion of the Sun
through the Local Bubble, for a plane passing through a meridian,
perpendicular to the galactic plane, and aligned with an axis
coinciding with the solar apex motion, extending from \glong=40\deeg\
to 220\deeg.  Stars with longitudes within $\pm$25\deeg\ of this
meridian slice are plotted.  Cleaned and averaged photometric and
astrometric data for O, B, and A stars in the Hipparcos catalog are
used, with variable stars and stars with poorly defined spectral types
omitted \cite{Frischetal:2006soljourn}.

In tenuous (``intercloud'') material, color excess, \ebv, and \NHI\
are related by \logNH/\ebv$>$21.70 (dex
\cmtwo\ mag$^{-1}$, \cite{BohlinSavageDrake:1978}).  The CLIC reddening is
\ebv$<$0.002 mag, and would not appear on this figure.  The large
circles show dust clouds from Dutra and Bica (2002). It is evident
that for a solar LSR velocity of $13-20$ pc
Myrs$^{-1}$, the Sun has been within the very low density Local Bubble
for over 3 Myrs.  Although CLIC column densities ($< 10^{19}$ \cmtwo)
are not traced by typical \ebv\ data, ultraviolet absorption lines do
not show any neutral ISM in the anti-apex direction,
\glong$\sim$220\deeg, at the \NHI$\sim 10^{17}$ \cmtwo\ level between
$\sim$5 and $\sim$100 pc (\cite{FrischYork:1983}, SF06).

\subsection{Local Bubble Interior} \label{sec:lbplasma}

We now discuss the properties of the paleolism when the Sun was in the very
low density LB interior.  The nature and origin of the Local Bubble is the
subject of ongoing debate.  We expect that this volume is presently filled
with a very low density ionized plasma that provides pressure support for the
cavity.  Observations of the diffuse soft X-ray background (SXRB) led to
models of the Local Bubble filled with hot, high pressure gas with a
temperature of $\sim 10^6$ K and pressure of $P/k_\mathrm{B} \sim 10^4$
cm$^{-3}$~K ($k_\mathrm{B}$ is the Boltzman constant), although ROSAT has
shown that substantial amounts of the SXRB at high latitude arises beyond the
boundaries of the Local Bubble \cite{McCammon:1983,KuntzSnowden:2000}.
An effective plasma temperature near $10^6$ K is indicated for a plasma in
collisional ionization equilibrium (CIE).  Disregarding the problems with
models of the emission spectrum, under the assumption that there is a CIE hot
plasma filling the LB, the implied density is roughly $5\times 10^{-3}$ \cc\
and pressure of $\sim P/k_\mathrm{B} = 10^4$ \cc K.

There are several implications of the SXRB attributed to the Local
Bubble.  First, the intensity of the background implies a thermal pressure 
that is substantially higher than that of the CLIC. This mismatch may be
fixed by a CLIC magnetic field of $B \sim 4$ \muG.  Second, a potentially
substantial contribution ($<$30\% in most directions) to the SXRB may arise
from charge transfer between solar wind ions such as O$^{+7}$ and neutral H or
He, either in the interstellar wind or geocorona
\cite{Cravens:2000,Cravensetal:2001}. This does not
affect the photoionization calculations described above directly, however,
even if the heliospheric and geocoronal soft X-ray emission is at the upper
end of current estimates.  That is because the heliospheric and geocoronal
soft X-ray emission is substantially harder than the emission that is directly
responsible for LIC ionization ($\sim 13.6 - 54.4$ eV).  There is an indirect
effect, though, insofar as the heliospheric X-ray emission affects our
understanding of the source of the SXRB, which in turn has implications for
the lower energy emission and the cloud interface.

This low density hot LB plasma constituted the solar galactic environment
for several million years.  The LB plasma properties have probably
been constant since then, because cooling times for low density
plasmas are over $\approx$100 Myrs.  Nevertheless, the LB properties
are not constant.  Interstellar shocks move through space at
velocities over 100 \kms\ ($\sim$100 pc Myrs$^{-1}$), quickly
traversing low density ISM.  Also, 25\% of the ISM mass is contained
in clouds traveling faster than 10 \kms\ through the LSR (based
on \HI\ 21-cm data in \cite{HTI}).  Therefore, over the past several Myrs
some dilution of the plasma must have occurred, particularly near the
LB boundaries.  The CLIC, with $V_\mathrm{LSR} \sim$18 \kms\ may be an
example.

The paleoheliosphere formed by the interactions between the very low
density plasma of the Local Bubble void and the solar wind has been
modeled by Mueller et al. (2006).  The resulting heliosphere
dimensions are similar to the present heliosphere, but the heliosheath
is thicker and pickup ions and other derivatives of interstellar
neutrals are absent.  These changes have opposite effects on the
modulation of galactic cosmic ray fluxes at the Earth.

\subsection{Interface between Local Bubble Plasma and CLIC}
\label{sec:interface}

One additional consequence of the presence of the Local Bubble plasma
is that a transition region must exist between the tepid CLIC gas and
the hot gas.  This transition ISM constituted the paleoLISM for a very
brief period of time.  Models of the interface as a thermally
conductive interface indicate a thin region of intermediate
temperature gas with the temperature changing by an order of magnitude
over a distance of $\sim 1000$ AU and more than doubling in the space
of 100 AU (\cite{Slavin:1989},SF02).  The temperature profile of
such a region is very steep near the cloud and flattens farther out as
the temperature approaches that of the hot gas. The density profile is
similarly shaped, though inverted because the thermal pressure remains
nearly constant in the interface.  The density increases into the
cloud by an order of magnitude over a distance of roughly 1000 AU.
Over this same distance the ionization is decreasing from nearly
complete ionization of H to the $\sim 30$\% of the cloud interior.

The cloud gas is evaporated by the influx of heat and flows off the
cloud, accelerated by a gentle pressure gradient created by the
thermal conduction.  If such a profile exists at the edge of the CLIC
gas, the Sun would have traversed from gas at $T\sim 10^5$ K to $\sim
7000$ K in a period of about 500 years.  This would result in a sudden
change in the heliospheric boundary conditions that could be very
disruptive and result in conditions far from any equilibrium
configuration.

An alternative possibility for the boundary is that of a turbulent
mixing layer \cite{Slavinetal:1993}.  This would be the case if there
is a substantial velocity difference between the CLIC gas and that of
the hot gas.  Even relative velocities on the order of the sound speed
in the hot gas or less could be very disruptive to the cooler clouds
and could lead to an interface in which the cool gas is being
entrained in the hotter gas, mixed and then cooled.  Such an interface
is similar in some ways to the evaporative boundary described above,
but depends on hydrodynamic instabilities to create the mixing of the
cloud gas into the hot bubble gas.  The crossing of this type of
interface could also be very disruptive of the heliosphere with the
likelihood of small scale condensations and velocity fluctuations as
well as sudden variations in ionization in the matter incident on the
Solar System.

\section{Discussion} \label{sec:assumptions} 

Simple assumptions are necessary to estimate the eras of the solar
transition between galactic environments, although we know that in 
general the ISM contains a wide range of cloud types and shows coherent 
structures ranging in size from $<$1 pc (e.g. the LIC) to $>$100 pc 
(e.g. Loop I).  Another limitation, following from the spectral analysis 
techniques used to acquire data, is that the identification of 
an interstellar ``cloud''
rests on a median velocity determined from the Doppler spread of
atomic velocities, which is highly sensitive to the
resolution of the instrument conducting the observations
\cite{WeltyNa:1994}.  

Over the ordinary long sightlines of the ISM, e.g. $>$100 pc,
similar simple assumptions are used for clouds that are 
obviously blended in velocity.  The more confined CLIC gas within $\sim$10 pc
allows higher levels of accuracy.  Howevever, although the
electron density diagnostics \MgI/\MgII\ and \CIIstar/\CII show
consistent values of \nel$\sim$0.1 \cc\ towards several stars sampling
the CLIC, the neutral
densities must be reconstructed from radiative transfer models because
\CI\ fine-structure data are unavailable \cite{JenkinsTripp:2001}.  
Model results indicate consistently that the CLIC does not fill the
sightline towards any nearby star, with filling factors of $\sim$0.40
in the galactic center hemisphere and $\sim$0.26 in the anti-center
hemisphere for \nHI$\sim$0.2 \cc\ \cite{FrischSlavin:2006}.  Except
for observations of \SiIII\ towards local stars
\cite{GryJenkins:2001} and pickup ion Ne, radiative transfer models
successfully reproduce densities of ISM towards $\epsilon$ CMa and
within the solar system.  Therefore, except for possible hidden
dense small clumps of gas, we expect that a cloud density of \nHI$\sim$0.2 \cc\
is a reasonable value.  Less certain is whether there are gaps in the
distribution of the nearest ISM, as would be expected for filamentary
LIC gas \cite{Frisch:1994}.

Nevertheless, our understanding of the nearest ISM is not complete,
and in particular the origin of the observed macroturbulence indicated
by the velocity data \cite{FGW:2002} and the possibility of denser
material in the upwind direction \cite{Frisch:2003apex} indicate much
of the local ISM remains a mystery.  The LIC is our best understood
cloud, yet distinctly different estimates of the date of the Sun's
entry in the LIC are found depending on which subset of data are used
to define an assumed filamentary LIC shape (resulting in entry
eras of less than $\sim$ 10,000 years ago and up to $\sim$40,000 years
ago, \cite{Frisch:1994,Muelleretal:2006}). Our analysis here of the paleoLISM
provides a starting point for future studies based on improved data
and a deeper understanding of turbulence and small-scale structure in
the CLIC.

The contribution to the soft X-ray background (SXRB)
radiation field at the Sun due to charge exchange between the
solar-wind and interstellar neutrals was discussed in \S \ref{sec:lbplasma}.
The primary effect of this emission will be on the properties of the
cloud interface, while the overall soft X-ray flux will continue
to be dominated by the strong emission from the Loop I supernova
remnant and attenuated extragalactic sources.

\input{6001-t01}

\section{Conclusions}

The primary conclusion of this paper is that, over the past several
million years, both the galactic environment of the Sun and the
heliosphere have been significantly different than they are today.
Observational data combined with theoretical studies can be used to
reconstruct the three-dimensional distribution of nearby ISM, and
predict the times the Sun transitioned between different environments.
If we assume a continuously distributed local ISM, within the past
$\sim 130,000 \pm$70,000 years, and possibly as recent as $\sim$56,000
years ago, the Sun entered low density partially ionized ISM flowing
away from the direction of the Scorpius-Centaurus Association.  Sometime within
the past $\sim$40,000 years the Sun entered the cloud now surrounding
the solar system, the LIC.  These estimates rely on topologically
simple models of the cluster of local interstellar clouds (CLIC)
flowing past the Sun; more elaborate models are discussed elsewhere
(\cite{Frisch:1994,Gry:1996,Muelleretal:2006}, SF06).  As the Sun
moves through this complex of local interstellar clouds, the boundary
conditions of the heliosphere should change by substantial amounts due
to changes in cloud temperature, velocity, and opacity-driven
variations in the ionization of the surrounding ISM.  Prior to that,
the Sun was in the low density plasma of the Local Bubble cavity.
Between the Local Bubble cavity and the CLIC, the Sun briefly
($\sim$500 years) passed through an interface region of some type.

These estimates of the entry date of the Sun into the CLIC and LIC
are based on
current data and models.  Future very-high
resolution observations in the 1000--3000 \AA\ spectral interval, for
a spatially dense sample of nearby stars, are required to reconstruct
the distribution, kinematics, and properties of the CLIC, and 
reduce the uncertainties in these estimates.  When that happens,
we can anticipate that the early work of Hans Fahr will have yielded 
grand results, as we finally understand the close
relationship between the paleoheliosphere and paleoLISM.


{\bf Acknowledgements:}
The authors acknowledge support for this research by NASA grants 
NAG5-11005, NAG5-13107, NAG5-13558, and \\ NNG05GD36G.


\vspace{0.1in}
Note:  To be published in Astrophysics and Space Sciences
Transactions (ASTRA), for the proceedings of the workshop "Future
Perspectives in Heliospheric Research: Unsolved Problems, New Missions
- New Sciences" Bad Honnef, Germany, April 6-8, 2005, held in honor of
Prof. Hans Fahr's 65th birthday.

\end{document}

%% file: 6001-journallist.tex
\newcommand\aap{{A\&A}}%
\newcommand\aj{{AJ}}%
\newcommand\apjl{{ApJ}}%
\newcommand\apjs{{ApJS}}%
\newcommand\apj{{ApJ}}%
\newcommand\apss{{Ap\&SS}}%
\newcommand\grl{{Geophys.~Res.~Lett.}}%
\newcommand\jgr{{J.~Geophys.~Res.}}%
\newcommand\jatp{{J.~Atmos.~Terr.~Phys.}}%
\newcommand\memras{{MmRAS}}%
\newcommand\mnras{{MNRAS}}%
\newcommand\nat{{Nature}}%
\newcommand\pasp{{PASP}}%
\newcommand\ssr{{Space~Sci.~Rev.}}%

%% file: 6001-t01.tex
\begin{table}[t]
   \caption{HC and LSR Velocities of the Sun, CLIC, and LIC}
  \vskip4mm
  \begin{center}
    \begin{tabular}{lc c c }
      \hline
& Sun  &  Upwind & Upwind \\
& Motion &  CLIC & LIC \\
      \hline
HC: &&\\
V (\kms) & -- & --28.1$\pm$4.6 & --26.3$\pm$0.4 \\
\vspace*{0.1cm} \glong,\glat & --  & 12.4\deeg, 11.6\deeg & 3.3\deeg, 15.9\deeg \\
LSR$_\mathrm{Std}$: &&\\
\vspace*{0.1cm} V (\kms) &  19.5 & --19.4 & --20.7 \\
\vspace*{0.1cm} \glong,\glat & 56\deeg, 23\deeg & 331.0\deeg,--5.1\deeg & 317.8\deeg, --0.5\deeg \\
LSR$_\mathrm{Hip}$: &&\\
\vspace*{0.1cm} V (\kms) &  13.4 & --17.0 & --15.7 \\
\vspace*{0.1cm} \glong,\glat & 27.7\deeg, 32.4\deeg & 2.3\deeg, --5.1\deeg & 346.0\deeg, 0.1\deeg\\ 
      \hline
    \end{tabular}
  \end{center}
Note:  The Standard solar motion, LSR$_\mathrm{Std}$, corresponds 
to a velocity of 19.5 \kms\ towards \glong=56\deeg, \glat=23\deeg.  
The Hipparcos solar motion, LSR$_\mathrm{Hip}$, corresponds to a 
velocity of 13.4 \kms\ towards \glong=27.7\deeg, \glat=32.4\deeg\ 
\cite{DehnenBinney:1998}. 
\label{tab:1}
\end{table}